\documentclass[aps,prx,reprint, superscriptaddress]{revtex4-1} 

\usepackage{soul}
\usepackage{lineno}
\usepackage{graphicx}
\usepackage{dcolumn}   
\usepackage{bm}        
\usepackage{amssymb}   
\usepackage{dsfont}
\usepackage{relsize}
\usepackage{amsmath,stmaryrd}
\usepackage{blkarray, multirow, graphicx, diagbox, color, xcolor, colortbl}
\usepackage{bbm, bbold}
\usepackage{glossaries}
\usepackage{hyphenat}
\usepackage{ifthen}
\usepackage[colorlinks, linkcolor = blue, citecolor = blue, filecolor = black, urlcolor = blue]{hyperref}
\usepackage{xkeyval}
\usepackage{moreverb}
\usepackage{rotating}
\usepackage{wrapfig}
\usepackage{slashbox}
\usepackage{xspace}
\usepackage{nicefrac}
\usepackage[]{units}
\usepackage{physics}
\usepackage{booktabs}
\usepackage{braket}
\usepackage[inline]{enumitem}
\usepackage{tabto}
\usepackage{listings}
\usepackage{xstring}
\usepackage{url}
\def\ReplaceStr#1{%
	\IfSubStr{#1}{p}{%
		\StrSubstitute{#1}{p}{.}}{#1}}

\usepackage[caption=false]{subfig}
\usepackage[capitalise]{cleveref} %
\usepackage{chemformula}
\usepackage{algorithm}

\newcommand{\nodagger}[0]{{\vphantom{\dagger}}}

\newcommand{\discussive}[1]{\textcolor{black}{#1}}

\usepackage{tikz}
\usetikzlibrary{calc}

\usepackage{xcolor}
\definecolor{mygreen}{HTML}{009000}
\definecolor{myorange}{HTML}{ff6600}
\definecolor{mypurple}{HTML}{6F00FF}
\definecolor{electriclime}{rgb}{0.8, 1.0, 0.0}

\definecolor{block}{RGB}{0,162,232}

\newenvironment{blockmatrix}{%
  \left(%
  \vcenter\bgroup\hbox\bgroup
  \tikzpicture[
    x=1.5\baselineskip,
    y=1.5\baselineskip,
  ]%
}{%
  \endtikzpicture
  \egroup
  \egroup
  \right)%
}

\newcommand*{\block}[1][block]{%
  \blockaux{#1}%
}
\def\blockaux#1(#2,#3)#4(#5,#6){%
  \draw[fill={#1}, draw=none]
  let \p1=(#2,#3),
      \p2=(#5,#6),
      \p3=(#2+#5,#3+#6),
      \p4=(#2+#5/2,#3+#6/2)
  in
    (\p1) rectangle (\p3)
    (\p4) node {\centering $#4$}
  ;%
}

\newacronym{OBC}{OBC}{open boundary conditions}
\newacronym{TFIM}{TFIM}{transverse-field Ising model}
\newacronym{KMS}{KMS}{Kubo-Martin-Schwinger}
\usepackage[section]{placeins}

\graphicspath{{../figures_new/}}
\begin{document}
\def\thetitle{Thermodynamics of the quantum Mpemba effect} 

\author{Mattia Moroder}
\email{mattia.moroder@physik.uni-muenchen.de}
\affiliation{Department of Physics, Arnold Sommerfeld Center for Theoretical Physics (ASC),
Munich Center for Quantum Science and Technology (MCQST),
Ludwig-Maximilians-Universit\"{a}t M\"{u}nchen, 80333 M\"{u}nchen, Germany.}
\author{Oisín Culhane}
\email{oculhane@tcd.ie}
\affiliation{School of Physics, Trinity College Dublin, Dublin 2, Ireland}
\author{Krissia Zawadzki}
\email{krissia@ifsc.usp.br}
\affiliation{School of Physics, Trinity College Dublin, Dublin 2, Ireland}
\affiliation{Instituto de Física de São Carlos, Universidade de São Paulo,
CP 369, 13560-970 São Carlos, São Paulo, Brazil}
\author{John Goold}
\email{gooldj@tcd.ie}
\affiliation{School of Physics, Trinity College Dublin, Dublin 2, Ireland}
\affiliation{Trinity Quantum Alliance, Unit 16, Trinity Technology and Enterprise Centre, Pearse Street, Dublin 2, D02YN67}

\title{\thetitle}
\begin{abstract}
We investigate the quantum Mpemba effect from the perspective of non-equilibrium quantum thermodynamics by studying relaxation dynamics \discussive{of quantum systems coupled to a Markovian heat bath}, which are described by Davies maps. Starting from a state with coherences in the energy eigenbasis, we demonstrate that an exponential speedup to equilibrium will always occur if the state is transformed to a diagonal state in the energy eigenbasis, provided that the spectral gap of the generator is defined by a complex eigenvalue. When the transformed state has a higher non-equilibrium free energy, we argue using thermodynamic reasoning that this is a \textit{genuine} quantum Mpemba effect. Furthermore, we show how a unitary transformation on an initial state can always be constructed to yield the effect and demonstrate our findings by studying the dynamics of both the non-equilibrium free energy and the irreversible entropy production in single and multi-qubit examples. 
\end{abstract}
\maketitle
The Mpemba effect describes a situation where an initially hot system is quenched into a cold bath and reaches equilibrium faster than an initially cooler system. Although the first systematic investigations of this phenomenon were performed by Mpemba, Osborne, and Kell~\cite{mpemba_original,Kell_freezing} in the late sixties, it had been discussed by Aristotle over 2000 years ago~\cite{Aristotle} and noticed by others, such as Descartes~\cite{Descartes1986-lv} and Bacon~\cite{Bacon1854-BACBNO}, throughout history. Since Mpemba and Osborne's original work, the effect has been explored in an increasingly diverse range of physical systems such as clathrate hydrates~\cite{Ahn2016-sa}, polymers~\cite{mpemba_polymers}, magnetic alloys~\cite{mpemba_magnets}, carbon nanotube resonators~\cite{Greaney2011-iq}, granular gases~\cite{mpemba_granular_fluids} and dilute atomic gases~\cite{Keller_2018}. However, despite the wide range of studies, the physical origin and even the very existence of the phenomenon are still debated in the literature~\cite{Burridge2016-qf,Bechhoefer2021-ah}.

A breakthrough in the understanding of this anomalous effect came from analysing the stochastic thermodynamics of Markovian systems~\cite{mpemba_theory, PhysRevX.9.021060}. At long times, the dynamical evolution of the state is a linear combination of the stationary state and the system's slowest decaying mode. A Mpemba effect can occur if the initial state with higher temperature has a smaller amplitude with this mode, with an exponential speedup to equilibrium occurring if the amplitude goes to zero (known as the strong Mpemba effect)~\cite{PhysRevX.9.021060}. This framework has been used recently to experimentally probe the Mpemba effect~\cite{exp_mpemba_nature_colloidal} and its inverse~\cite{kumar2022anomalous} under controlled conditions and to boost thermalisation protocols~\cite{PhysRevLett.124.060602}.

\begin{figure}
    \centering
    \includegraphics[width=0.655\columnwidth, trim={1mm 2mm 2mm 2mm}]{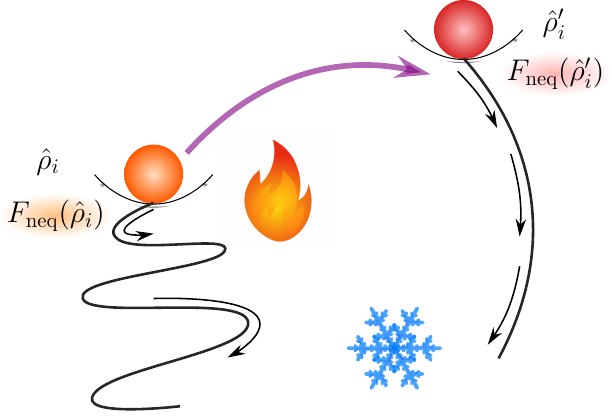}
    \caption{
 For a general quantum state, one can construct a unitary transformation, which increases the state's non-equilibrium free energy and eliminates the overlap with a set of slow-decaying modes of the generator. This leads to exponentially faster thermalisation and a genuine quantum Mpemba effect. }
    \label{fig:mpemba_cartoon}
\end{figure}
Recently work has been undertaken to generalise the Mpemba effect to quantum dynamics. The phenomenon and the framework generalise naturally to open quantum systems~\cite{PhysRevB.100.125102, carollo_prl, bao2022accelerating, PhysRevResearch.3.043108, carollo_numerical, PhysRevE.108.014130, PhysRevLett.131.080402, chatterjee2023multiple, wang2024mpemba, caceffo2024entangled, strachan2024nonmarkovian} and this has lately been explored experimentally in ion traps~\cite{shapira2024mpemba, strong_mpemba_observation}. 
However, it should be stated that very few of these studies focus specifically on thermalisation but rather more generally on anomalous relaxation to general fixed points.
Another series of works have investigated a quantum Mpemba-like effect in isolated systems, related to symmetry restoration in quenched dynamics starting from states that break the symmetry of the generating Hamiltonian~\cite{mpemba_nature_calabrese, calabrese_mpemba_xy_chain, yamashika2024entanglement, liu2024symmetry,joshi2024observing}.

In this Letter, we focus specifically on Davies maps which are \discussive{an important class} of continuous-time quantum dynamical semigroups \discussive{describing thermalisation}.
These maps rigorously describe the thermalisation of a quantum system when weakly coupled to a heat bath~\cite{DAVIES1979421,ROGA2010311,PhysRevResearch.3.023006} and are often said to be the quantum equivalent of classical Glauber dynamics.
The special mathematical properties of these maps allow us to identify a unique quantum Mpemba effect that stems from the subblock structure of the generator in the energy eigenbasis.
Starting from a state that has coherences, we show how a unitary transformation can transform the state to generate an exponential speed-up which will always occur when the spectral gap of the generator is defined by a complex eigenpair.
When the transformed state has a higher non-equilibrium free energy~\cite{donald1987free,esposito2011second,parrondo2015thermodynamics} the dynamics give rise to a genuine quantum Mpemba effect (see~\cref{fig:mpemba_cartoon}).
Furthermore, we study the division of the irreversible entropy production~\cite{10.1063/1.523789,RevModPhys.93.035008} into coherent and incoherent parts~\cite{Santos2019-vt, PhysRevE.99.042105}.
Our findings are illustrated on single-qubit and many-qubit examples.   

Consider a quantum Markovian master equation that evolves a density matrix $\hat{\rho}(t)$ 
\begin{equation}
\frac{d\hat{\rho}(t)}{dt}=\mathcal{G}\hat{\rho}(t).
\end{equation}
Here $\mathcal{G}$ is the generator, known as the Lindbladian, consisting of a unitary part and a dissipative part such that $\mathcal{G}=i\delta+\mathcal{D}$ with $\delta(\mathord{\cdot})=-[\hat{H},\mathord{\cdot}]$, $\hat{H}$ being the system Hamiltonian and $\mathcal{D}(\mathord{\cdot})=\sum_l \hat{L}^{\nodagger}_l(\mathord{\cdot}) \hat{L}^{\dagger}_l -\frac{1}{2} \{ \hat{L}^{\dagger}_l\hat{L}^{\nodagger}_l, (\mathord{\cdot}) \}$, where $\hat{L}_{k}$ are the so-called jump operators. Using the generator we can write down the general evolution of an initial density matrix $\hat{\rho}_{i}$ as 
\begin{equation}
\label{eq:generator}
\hat{\rho}(t)=e^{\mathcal{G}t}\hat{\rho}_i=\hat{\tau}+\sum_{k=2}^{D^2}\Tr(\hat{l}_{k}\hat{\rho}_i)\hat{r}_ke^{\lambda_k t} \:,
\end{equation}
where $D = \mathrm{dim}(\hat{H})$ and we assume that $\hat{\tau}$ is the unique steady state that is the right eigenoperator of the generator $\mathcal{G}$ associated with the zero eigenvalue. Here $\hat{l}_{k}$ and $\hat{r}_{k}$ are the left and right eigenoperators corresponding to the eigenvalue $\lambda_{k}$ such that 
$\mathcal{G}[\hat{r}_{k}]=\lambda_{k}\hat{r}_{k}$ and $\mathcal{G}^{\dagger}[\hat{l}_{k}]=\lambda_{k}\hat{l}_{k}$ where $\mathcal{G}^{\dagger}$ is the adjoint generator that acts on observables rather than states. The superoperator $\mathcal{G}$ preserves the Hermiticity of $\hat{\rho}(t)$, which implies that if $\lambda_k$ is a complex eigenvalue then $\lambda^{*}_{k}$ is also. We order the eigenvalues in ascending order according to the modulus of their real part such that $0=\lambda_{1}<|\Re(\lambda_{2})|\le|\Re(\lambda_{3})|\le\dots$. The spectral gap is then $\abs{\Re(\lambda_{2})}$ and defines the longest timescale in the system such that $|\hat{\rho}(t)-\hat{\tau}|\propto \exp(\Re(\lambda_{2})t)$. In~\cite{carollo_prl} the authors show that an exponential speed up to the steady state can be achieved by finding a unitary that acts on the initial state such that 
\begin{equation}
\Tr(\hat{l}_{2}\hat{U} \hat{\rho}_{i}\hat{U}^{\dagger})=0,
\end{equation}
so that $|\hat{\rho}(t)-\hat{\tau}|\propto \exp(\Re(\lambda_{3})t)$. The key assumptions in~\cite{carollo_prl} were a pure initial state and a spectral gap defined by a real eigenvalue. When the lowest eigenmodes of the generator form a complex conjugate pair at long times one has
\begin{equation}
\label{eq:coherences}
\begin{split}
    \hat{\rho}(t) \propto & \;\hat{\tau}+e^{\Re(\lambda_2)t} \times \\ &\left ( \Tr(\hat{l}_{2}\hat{\rho}_i)\hat{r}_2e^{i\Im(\lambda_2)t}+\Tr(\hat{l}^\dagger_{2}\hat{\rho}_i)\hat{r}_2^{\dagger} e^{-i\Im(\lambda_2)t} \right),
\end{split}
\end{equation}
so for a strong Mpemba effect, one would need to find a unitary such that both $\Tr(\hat{l}^\dagger_{2}\hat{U}\hat{\rho}_{i}\hat{U}^{\dagger})=0$ and $\Tr(\hat{l}_{2}\hat{U}\hat{\rho}_{i}\hat{U}^{\dagger})=0$~\cite{carollo_numerical}. 
As argued in~\cite{carollo_numerical}, $\hat{\l}_{2}$ is non-Hermitian so the strong Mpemba effect cannot be found using the same logic as in~\cite{carollo_prl}.
Here we demonstrate an alternative route to obtain such a unitary which exploits the mathematical structure of the generator in thermalising open systems.

We consider Davies maps, which are defined by two properties. First, the unitary and dissipative part of the generator commute, and second they obey quantum detailed balance~\cite{alicki1976detailed} with respect to the thermal state $\hat{\tau}_{\beta}=e^{-\beta\hat{H}}/Z$, with $Z = \text{Tr}\left(\text{exp}\left(-\beta\hat{H}\right)\right)$. Mathematically quantum detailed balance means that given two arbitrary operators $\hat{A}$ and $\hat{B}$, the unitary and dissipative parts of the generator obey $\langle \hat{A}, \mathcal{D}^{\dagger}(\hat{B})\rangle_{\hat{\tau}_{\beta}}=\langle \mathcal{D}^{\dagger}(\hat{A}),\hat{B}\rangle_{\hat{\tau}_{\beta}}$ 
and $[\hat{H},\hat{\tau}_{\beta}]=0$,
where $\langle\hat{A},\hat{B}\rangle_{\hat{\tau}_{\beta}}=\Tr(\hat{\tau}_{\beta}\hat{A}^{\dagger}\hat{B})$ is the \gls{KMS} inner product. The eigenmatrices of the unitary part $\delta$ are the elemental matrix units $|n\rangle\langle m|$ with eigenvalues $\omega_{nm}=E_{n}-E_{m}$. Since this is a Davies map, the dissipative part $\mathcal{D}$ shares a common eigenspace with $\delta$. Furthermore, if the Hamiltonian is non-degenerate, then the overall Davies map $e^{\mathcal{G}t}$ is block diagonal in the energy eigenbasis. We can write $\mathcal{G}=\mathcal{G}_{P}\mathop{\oplus}\mathcal{G}_{C}$, where $\mathcal{G}_{P}=\mathcal{D}_{P}$ is the population subblock, whose right eigenmatrices are diagonal in the energy eigenbasis $|n\rangle\langle n|$ and $\mathcal{G_{C}}=i\delta+\mathcal{D}_C$ is the coherence subblock. Note that due to the detailed balance condition the eigenvalues corresponding to the off-diagonal eigenmatrices of $\mathcal{D}_C$ are real whereas imaginary eigenvalues are due to the unitary part $\delta$. The slowest decaying modes shown in~\cref{eq:coherences} and indeed more generally $\hat{r}_i$ for $i\ge 2$ will be off-diagonal matrices in the energy eigenbasis. This ensures that they are traceless, which guarantees that the dynamics preserve normalisation at all times. However, not only does $\delta$ commute with $\mathcal{D}$, it also commutes with $\mathcal{D}^{\dagger}$. This means that the left eigenvectors of $\mathcal{G_{C}}$ are also purely off-diagonal matrices in the energy eigenbasis since $\mathcal{G}^{\dagger}[\hat{l}_{k}]=\lambda_{k}\hat{l}_{k}$. 

We now state our first finding: given an initial state $\hat{\rho}_i$ which has coherences in the energy eigenbasis, under an evolution with a Davies generator which has a spectral gap defined by the real part of a complex eigenpair, an exponential speed up towards the fixed point can always be achieved if a unitary $\hat{U}$ is performed to bring the state to be diagonal in the energy eigenbasis. Since all $\hat{l}_m$ with $\lambda_m\in \mathbb{C}$ are purely off-diagonal matrices in the energy eigenbasis then all associated amplitudes will be simultaneously eliminated $\Tr(\hat{l}_m\hat{U}\hat{\rho}_{i}\hat{U}^\dagger)=0$.
\begin{figure}
    \centering
\includegraphics[width=0.425\textwidth]{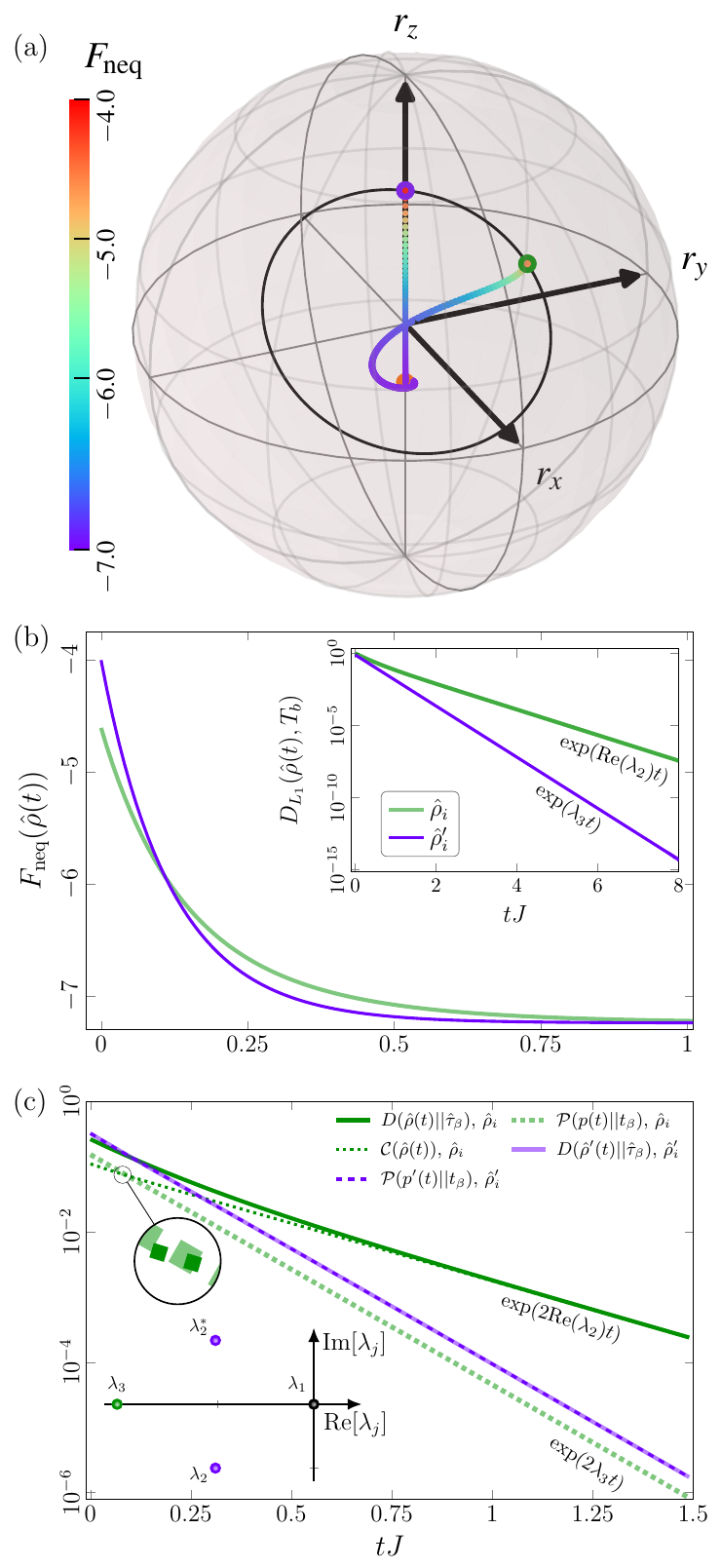}
    \caption{
 The genuine quantum Mpemba effect in a single-qubit system. a) Bloch sphere representation of the equilibration of a random state $\hat{\rho}_i$ (green dot) and the transformed state $\hat{\rho}_i' = \hat{U} \hat{\rho}_i \hat{U}^\dagger$ (purple dot) towards the thermal steady state (orange dot). The non-equilibrium free energy  $F_{\mathrm{neq}}$ for the two thermalisation processes is represented by the color scheme. b) $F_{\mathrm{neq}}$ is initially higher for the optimized state (purple line), but drops below the random state's one (green line) at a later time. This crossing shows the occurrence of a genuine quantum Mpemba effect. \textit{Inset}: The $L_1$-distances from the steady state indicate that $\hat{\rho}_i'$ thermalises exponentially faster than $\hat{\rho}_i$.
 c) Dynamics of the total entropy production $D$ and its classical $\mathcal{P}$ and coherent $\mathcal{C}$ contributions. The rotation $\hat{U}$ yields an exponential speed up. \emph{Inset}: spectrum of the Davies generator $\mathcal{G}$ for a single qubit.
 Here $\omega=5J$, \discussive{$\boldsymbol{r}_i= (0.276, 0.359, 0.303)$}, $\boldsymbol{r}_i'= (0., 0., 0.545)$, and $T_b =10 J$.
 }
    \label{fig:single_qubit}
\end{figure}
We note that alternative protocols exploiting quantum coherences to speed up equilibration have been proposed in~\cite{PhysRevLett.122.250402,boubakour2024dynamical}.

We want to stress an essential point: an exponential speedup in itself does not constitute a {\it{genuine}} Mpemba effect. In the original breakthrough paper~\cite{mpemba_theory}, which focuses on thermodynamics, the effect is defined according to states with three temperatures $T_{b}<T_{c}<T_{h}$ which pertain to the fixed point, colder and hotter initial states, respectively. A Mpemba effect occurs if some time $t_m$ exists after which $D[{\hat{\rho}}^{h}(t),T_b]<D[\hat{\rho}^{c}(t),T_b]$. Here $D[\hat{\rho}(t),T]$ denotes a distance measure between the time-evolving state and the fixed point. Since we are dealing with non-equilibrium quantum states with possible coherences in the energy eigenbasis we do not have any initial notion of temperature. We have one point of equilibrium, namely, the fixed point $\hat{\tau}_{\beta}$. For quantum dynamics, we propose to use the non-equilibrium free energy which is defined as 
\begin{equation}
\label{eq:nf_E}
F_{\textrm{neq}}(\hat{\rho}(t))=\Tr(\hat{H}\hat{\rho}(t))+\frac{1}{\beta}\Tr(\hat{\rho}(t)\ln \hat{\rho}(t)) .
\end{equation}
We consider the initial scenario $F_{\textrm{neq}}(\hat{\rho}'(t_{0}))>F_{\textrm{neq}}(\hat{\rho}(t_0))>F_{\textrm{eq}}$ where $\hat{\rho}'(t_0)=\hat{U}\hat{\rho}(t_0)\hat{U}^{\dagger}$ is the state following unitary transformation and $F_{\textrm{eq}}=-\beta^{-1}\ln Z$ is the equilibrium free energy of the fixed point. A quantum Mpemba effect occurs if some time $t_{m}$ exists such that $F_{\textrm{neq}}(\hat{\rho}'(t))<F_{\textrm{neq}}(\hat{\rho}(t))$ for all times $t>t_m$. We call this a \textit{genuine} quantum Mpemba effect to differentiate it from exponential speedups achieved by other transformations. We emphasize that for the speedup to be a quantum Mpemba effect the non-equilibrium free energy curves need to cross in time. The non-equilibrium free energy has several additional features that make it ideal to define and analyse the quantum Mpemba effect. First of all, it can be written as 
\begin{equation}
    \label{eq:relative_e}
 F_{\textrm{neq}}(\hat{\rho}(t))=\beta^{-1}D(\hat{\rho}(t)||\hat{\tau}_{\beta})+F_{\textrm{eq}}(\hat{\tau}_{\beta}), 
\end{equation}
where $D(\hat{\rho}||\hat{\sigma})=\Tr[\hat{\rho}(\ln\hat{\rho}-\ln\hat{\sigma})]$ is the quantum relative entropy. The Klein inequality guarantees the positivity of the relative entropy, ensuring that $F_{\textrm{neq}}(\hat{\rho}(t))\ge F_{\textrm{eq}}(\hat{\tau}_{\beta})$ $\forall t$. The quantum relative entropy is a very stringent measure of the distinguishability of two quantum states. While not itself a metric, it still upper bounds the trace distance via Pinsker’s inequality $D(\hat{\rho}||\hat{\sigma})\ge \vert\lvert \hat{\rho}-\hat{\sigma}\rvert\rvert_{1}^{2}/2 $, which captures the optimal distinguishability of quantum states with a single measurement.
In addition,~\cref{eq:nf_E} gives us a prescription to find operations that transform an initial state to a state that has both a higher $F_{\text{neq}}$ and also is diagonal in the energy eigenbasis. A simple operation that does this is a unitary which diagonalises the state in the energy eigenbasis and then does a population inversion (see~\cref{app:transforming:inital:state}). When the generator has a spectral gap defined by a complex eigenpair this will {\it always} yield a quantum Mpemba effect since all overlaps with coherent modes go to zero. Finally, not only is the non-equilibrium free energy directly connected to the Spohn entropy production rate~\cite{10.1063/1.523789} as $\Pi=-\beta^{-1} dF_{\textrm{neq}}(\hat{\rho}(t))/dt$, but as shown in~\cite{Santos2019-vt, PhysRevE.99.042105} and in~\cref{app:Free energy}, the total irreversible entropy produced can be divided as
\begin{equation}
\label{eq:division}
 D(\hat{\rho}(t)||\hat{\tau}_{\beta})=\mathcal{P}(p(t)||\hat{\tau}_{\beta})+\mathcal{C}(\hat{\rho}(t)),
\end{equation}
where $\mathcal{P}(p(t)||t_{\beta})=\sum_{n}p_n(t)\ln\frac{p_n(t)}{t^{\beta}_n}$ is the classical relative entropy of the time-evolving population vector and the thermal state and $\mathcal{C}(\hat{\rho}(t))=S(\Delta(\hat{\rho}(t)))-S(\hat{\rho}(t))$ is known as the relative entropy of coherence where $\Delta(\hat{\rho}_{s}(t))$ is the state dephased in the systems' energy eigenbasis. This is known to be an operational measure of quantum coherence~\cite{plenio_rev}. 
In the following, we consider the thermalisation of single and multi-qubit systems. The characteristic energy scale is $J$ and we set $k_B=1$ and the dissipation strength $\gamma = J$ (see~\cref{app:Davies-Lindblad}).

\paragraph*{Example 1: Single Qubit} The qubit is prepared in a random state $\hat{\rho}_i = 1/2 (\mathds{1} + \boldsymbol{r} \cdot \hat{\boldsymbol{\sigma}} )$, where $\boldsymbol{r}$ is the Bloch vector and $\hat{\boldsymbol{\sigma}}$ is the vector containing the three Pauli matrices. 
The thermalisation is described by a unitary part governed by the Hamiltonian $\hat{H} =\omega/2 \hat{\sigma}_z$ and a dissipative part $\mathcal{D}$ defined according to the Davies map for a bosonic particle and a bath at temperature $T_b$.
We apply a unitary $\hat{U}$ on $\hat{\rho}_i$ which rotates the Bloch vector to $\boldsymbol{r}_i'$ and brings it to a new state $\hat{\rho}_i'$ diagonal in the energy eigenbasis with the eigenvalues in ascending order. 
The thermalisation dynamics of both states are pictured on the Bloch sphere in~\cref{fig:single_qubit}(a).
The color bar indicates the non-equilibrium free energy $F_{\mathrm{neq}}$, which is shown also in panel (b).
The crossing of the two lines indicates a genuine quantum Mpemba effect. 
The inset shows the $L_1$-distance from the steady state $D_{L_1}[\hat{\rho}(t), T_b] = \sum_{i,j} \abs{\hat{\rho}^{ij}(t) -\hat{\tau}_\beta^{ij} }$, that was studied also in the first experiment on classical Markovian systems~\cite{exp_mpemba_nature_colloidal} and indicates that the thermalisation for the transformed state $\hat{\rho}_i'$ is exponentially accelerated.
We now study the division of the entropy production defined by ~\cref{eq:division} in~\cref{fig:single_qubit}(c).
It can be seen that for the transformed state $\hat{\rho}_i'$ the coherent contribution is zero since the overlap with the coherent modes has been eliminated.
The spectrum of the Davies map is shown as an inset.
Note that for a single qubit, at any temperature $T_b$ the spectral gap is defined by a complex eigenpair.
\begin{figure}[htb!]
\includegraphics[width=0.95\columnwidth,trim={0.2cm 0 0 0},clip]{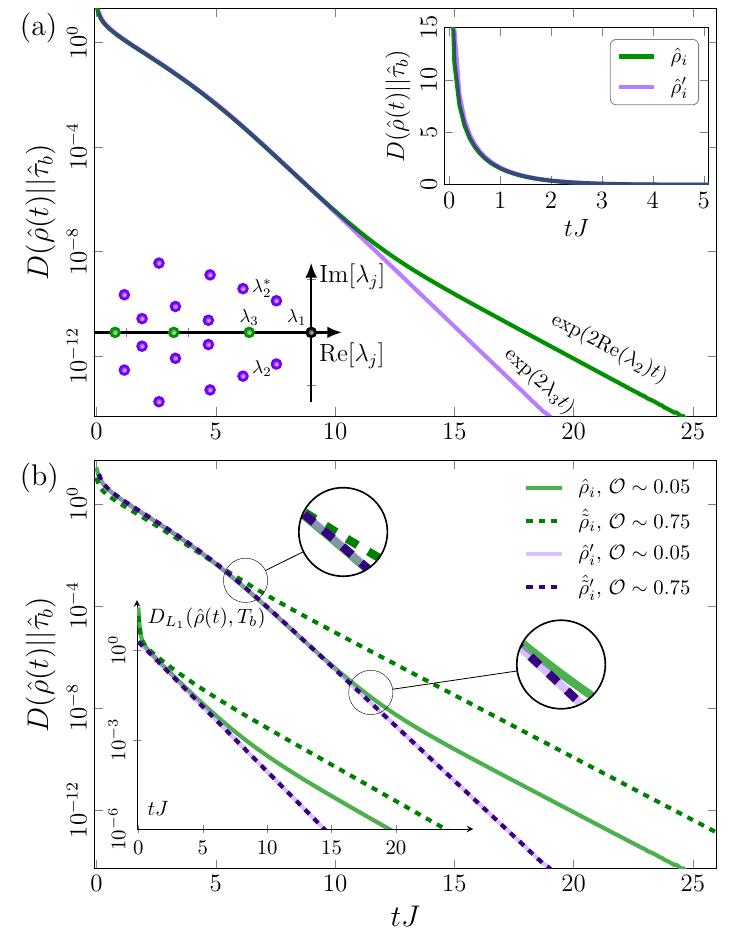}
\caption{
The genuine quantum Mpemba effect in a many-qubit system manifests at different timescales.
 (a) We study the total entropy production $D$ for the transverse field Ising model with anisotropy at $h=J/2$ with $L=5$ spins and bath temperature $T_b=0.1\,J$. 
 The rotation $\hat{U}$ allows for an exponential speedup governed by $\lambda_3$ that shows up at timescales much larger than for the single qubit case. 
 The upper inset shows $D$ in linear scale while the lower inset displays the first $20$ eigenvalues of the spectrum of the Davies generator $\mathcal{G}$.
 (b) An initial state with a large overlap with the two slowest decaying modes $\mathcal{O} = \abs{\Tr(\hat{l}^\dagger_2 \hat{\rho}_i)} + \abs{\Tr(\hat{l}_2 \hat{\rho}_i)}$ achieves the speedup sooner than one with a small overlap.
 For $\mathcal{O} \sim 0.05$ (full green line), the curves for the total entropy production cross around $t/J \sim 10$, while with $\mathcal{O} \sim 0.75$ (dashed green line), they coalesce at half of this time.}
\label{fig:mpemba:many:qubits}    
\end{figure}

\paragraph*{Example 2: Many Qubits}
We now shift our focus to a more complex system and consider a \gls{TFIM}  with open boundary conditions $\hat{H} = -J \sum_{j = 1}^{L-1} \hat{\sigma}^z_j \hat{\sigma}^z_{j+1} + h   \sum_{j = 1}^{L} \hat{\sigma}^x_j$.
We construct a random mixed state $\hat{\rho}_i$ by averaging over $1000$ random pure states. 
Then, as for the single qubit case, we obtain the transformed state $\hat{\rho}_i'$ by applying $\hat{U}$.
In panel (a) of~\cref{fig:mpemba:many:qubits} we compute the irreversible entropy production.
The main difference with the single qubit case is that since the overlap of the random state $\hat{\rho}_i$ with the slowest mode, $\hat{l}_2$, is small, the difference with the transformed state manifests only at a later time.
The situation changes when one considers an initial state $\hat{\tilde{\rho}}_i$ which has a larger overlap with $\hat{l}_2$.
We construct such a state by applying a Metropolis-based transformation, which is explained in~\cref{app:metropolis}. 
The dashed green line in panel (b) shows that similarly to the single qubit case, the exponential difference in the thermalisation speed of $\hat{\tilde{\rho}}_i$ and $\hat{\tilde{\rho}}_i'$ is evident already at short times. 
\discussive{ \paragraph*{Physical realisations}
For many physically relevant scenarios, starting from the microscopic Hamiltonian for system and environment and following the standard steps~\cite{BRE02}, one can derive an effective Davies map\hyp description.
These include, for instance, quantum optics setups~\cite{paule2018thermodynamics} and mesoscopic heat engines~\cite{Josefsson2018-me}.
Importantly, Davies maps can also be used to accurately estimate the impact of noise on single\hyp qubit and two\hyp qubit operations in superconducting transmon architectures~\cite{papič2023fastestimationphysicalerror}.
For concreteness, in~\cref{app:davies:applications}, we consider the Davies map\hyp description of a spinful fermionic quantum dot coupled to a heat bath and a two\hyp level atom in a photonic environment and reveal the occurrence of the Mpemba effect for experimental parameters.
Note also that the classical and quantum contributions to the entropy production(\cref{eq:division}) have recently been measured in a table\hyp top optical experiment described by a Davies map~\cite{xue2024evidence}.}
\paragraph*{Conclusion} We have demonstrated that a quantum Mpemba effect can be engineered by exploiting the subblock structure of the generator in thermalising open Markovian systems. We highlight the non-equilibrium free energy as a central thermodynamic object to both define and study the effect in open quantum systems. When the spectral gap of a Davies generator is defined by a complex eigenpair a unitary operation can always be found on an initial state that simultaneously raises the free energy and provides exponentially quicker thermalisation. 
The work presented here not only enhances our understanding of the effect in quantum systems and should inspire further studies from the perspective of quantum thermodynamics, but it helps bridge the gap between quantum and classical systems and could prove useful in dissipative state engineering protocols~\cite{Verstraete_2009_dissipative,Harrington_2022_dissipative,edo2024study}. 
\paragraph*{Acknowledgements} We are grateful to Mark T. Mitchison, Felix Binder, Federico Carollo, Gabriel Landi, Simon Milz, Laetitia Bettmann, and Alessandro Summer for discussions.  This work was supported by the EPSRC-SFI joint project QuamNESS. J.G. is supported by a SFI- Royal Society University Research Fellowship. O.C. is supported by the Irish Research Council under Grant No. 210500.
M.M. is supported by the Munich Center for Quantum Science and Technology.
\FloatBarrier
\bibliography{references}
\appendix
\section{\label{app:transforming:inital:state} Transforming the initial state}
In this appendix, we provide more details about the unitary transformation we perform on the initial state and prove that it always increases the non-equilibrium free energy.  

In the main text, we argued that to achieve asymptotically, exponentially faster relaxation when the spectral gap is defined by a complex eigenpair, we needed to perform a rotation such that the overlap with the slowest decaying modes go to zero i.e. $\Tr(\hat{l}_{2}\hat{U}\hat{\rho}_{i}\hat{U}^{\dagger})=0$ and $\Tr(\hat{l}^{\dagger}_{2}\hat{U}\hat{\rho}_{i}\hat{U}^{\dagger})=0$. We consider a unitary operator divided into two separate unitary operations
\begin{equation}
    \hat{U} = \hat{U}_2 \hat{U}_1,
\end{equation}
where $\hat{U}_1$ achieves the exponential, asymptotic speed up to relaxation and $\hat{U}_2$ ensures that the rotated state will have a higher non-equilibrium free energy than the initial state. The unitary $\hat{U}_1$ diagonalises the density matrix in the energy eigenbasis;
\begin{equation}
    \hat{U}_1 \hat{\rho}_i \hat{U}_1^\dag= \hat{\Lambda},
\end{equation}
where $\hat{\Lambda}$ is the diagonal matrix of eigenvalues of the density operator. This diagonalisation removes overlaps with all coherent modes, achieving an exponential speed-up if the spectral gap is defined by the real part of a complex eigenpair. We define $\hat{U}_2$ to swap the populations of the diagonalised density matrix, generating a population inversion. 

To get a genuine Mpemba effect we required our rotated state to have a higher non-equilibrium free energy $F_{\text{neq}}(\hat{\rho})$ than the initial state. We will now show that the rotation defined above will increase the non-equilibrium free energy of an arbitrary initial state $\hat{\rho}_i$. We need only compare the relative entropy of the states before and after the rotation, i.e.
\begin{equation}
 D(\hat{\rho}_i||\hat{\tau}_\beta) \leq D(\hat{\rho}'||\hat{\tau}_\beta).
\end{equation}

Since $\hat{U}$ is unitary, it does not change the von Neumann entropy of $\hat{\rho}_i$. With that, we need to show 
\begin{equation}
    \label{Eq: ineq}
    \text{Tr}[\hat{\rho}_i\text{ln}(\hat{\tau}_\beta)] \geq \text{Tr}[\hat{\rho}'\text{ln}(\hat{\tau}_\beta)].
\end{equation}
As the thermal state is diagonal in the energy eigenbasis and the populations are non-increasing, comparing the free energies is equivalent to comparing the populations before and after the rotation. 
The population distribution of the thermal equilibrium state is in decreasing order. Thus, to maximise $F_{\text{neq}}$, the populations of $\hat{\rho}'$ need to be ordered in increasing value, i.e. a complete population inversion. 

To demonstrate that the unitary $\hat{U}_2$ produces a state with the desired populations, we use the concept of majorisation~\cite{Bhatia_1997}. 
Denote $p^\downarrow$ to be the permutation of the distribution $p$ with values in descending order. One distribution $p$ is said to majorise another distribution $r$, $p\succ r$ if
\begin{equation}
    \sum_{j=1}^k p^\downarrow_j \geq \sum_{i=1}^k r^\downarrow_j \:\forall\: k.
\end{equation}
The probability distribution $p$ majorises $r$ if it is ``less spread out". 
As the thermal state populations in Eq.~\ref{Eq: ineq} are diagonal matrices with decreasing populations, they act as weights on the population probabilities of the density matrices. If the diagonalised state majorises any unitary transformation, then the full population invertion maximises the non-equilibrium free energy given the constraint of unitary operations. We are interested in any arbitrary unitary from the population-inverted state
\begin{equation}
    \sum_k\bra{k} \hat{U} \hat{\rho}' \hat{U}^\dag\ket{k} = p_\sigma,
\end{equation}
where $\ket{k}$ are the eigenvectors of the Hamiltonian and $p_\sigma$ are the populations of a unitarily connected density matrix. We expand $\hat{\rho}'$ into its eigenbasis
\begin{align}
    \sum_r \bra{k} \hat{U} \ket{r} p_r \bra{r} \hat{U}^\dag\ket{k} &= \sum_r |\bra{k} \hat{U} \ket{r}|^2p_r \nonumber \\
    &= \sum_r S_{kr}p_r = p_\sigma,
\end{align}
where $\hat{S}$ is a doubly stochastic matrix that maps the probability distribution $p_r$ to the probability distribution $p_\sigma$. From the Hardy-Littlewood-Polya inequality~\cite{Hardy_Littlewood_Polya_1934} such a doubly stochastic matrix exists if and only if the distribution $p_r$ majorises the distribution $p_\sigma$. Thus this final state has a higher non-equilibrium free energy than any other state connected by a unitary rotation with equality if and only if the initial state is diagonal in the energy eigenbasis with a full population inversion.

\section{\label{app:Free energy} Relative entropy and non-equilibrium free energy}
In this appendix, we show how the non-equilibrium free energy is written in terms of relative entropy. This is used extensively in quantum and classical stochastic thermodynamics~\cite{esposito2011second,parrondo2015thermodynamics} but to the authors' knowledge, the connection was first highlighted in the literature by Donald in the 80s \cite{donald1987free}.
Consider the internal energy $U$ of an arbitrary, possibly time-dependent, state $\hat{\rho}(t)$. We can always write the expression by introducing a thermal state $\hat{\tau}_{\beta}$ as 
\begin{equation}
    \begin{split}
 U(\hat{\rho}(t))&=-\beta^{-1}\Tr[\hat{\rho}(t)\log e^{-\beta\hat{H}}]\\
    &=-\beta^{-1}\Tr[\hat{\rho}(t)\log \frac{e^{-\beta\hat{H}}}{Z}]-\beta^{-1}\Tr(\hat{\rho}(t)\log{Z})\\
    &=-\beta^{-1}\Tr[\hat{\rho}(t)\log \hat{\tau}_{\beta}]-\beta^{-1}S(\hat{\rho}(t))+\beta^{-1}S(\hat{\rho}(t))\\ &+F_\text{eq}\\
    &=\beta^{-1} D(\hat{\rho}(t)||\hat{\tau}_{\beta})+\beta^{-1}S(\hat{\rho}(t)) +F_\text{eq},
    \end{split}
\end{equation}
where on the third line we have added and subtracted a von Neumann entropy term $S(\hat{\rho})=-\Tr(\hat{\rho}\log{\hat{\rho}})$ and we have introduced the relative entropy between the state and the thermal state. The non-equilibrium free energy is then computed from $F_{\text{neq}}(\hat{\rho}(t))=U(\hat{\rho}(t))-\beta^{-1}S(\hat{\rho}(t))$ and we can use the expression for the internal energy above so that 
\begin{equation}
 F_{\textrm{neq}}(\hat{\rho}(t))=\beta^{-1} D(\hat{\rho}(t)||\hat{\tau}_{\beta})+F_{\text{eq}}
\end{equation}
as used in the main text. When $\hat{\rho}(t)=\hat{\tau}_{\beta}$ the equilibrium free energy is recovered $F_{\text{eq}}=-\beta^{-1}\log{Z}$. For more on the physical interpretation of $F_{\text{neq}}$ see \cite{parrondo2015thermodynamics}. 
\section{\label{app:Davies-Lindblad} Explicit form of the Davies map}
For pedagogical purposes, in this section, we give some more details regarding Davies maps studied in the main text.  
For Markovian open quantum systems,  the dynamics of the system's density matrix $\hat{\rho}$ is governed by the Lindblad
equation
\begin{equation}
    \frac{d\hat{\rho}(t)}{dt}=\mathcal{G}\hat{\rho}(t) = -i [\hat{H}, \hat{\rho}(t)] +\sum_l \hat{L}^{\nodagger}_l\hat{\rho}(t) \hat{L}^{\dagger}_l -\frac{1}{2} \{ \hat{L}^{\dagger}_l\hat{L}^{\nodagger}_l, \hat{\rho}(t) \},
    \label{eq:lindblad}
\end{equation}
where $\mathcal{G}$ is the generator, $\hat{H}$ is the system's Hamiltonian, and $\hat{L}^{\nodagger}_l$ are the jump operators describing the influence of the environment on the system~\cite{manzano_lindblad_intro}. 
An important class of Lindbladians, known as Davies maps~\cite{DAVIES1979421}, describes the thermalisation of a quantum system when weakly coupled to a heat bath.
The fixed point of Davies maps is the thermal state at inverse temperature $\beta$
\begin{equation} 
    \hat{\tau}_{\beta} = \exp\left( -\beta \hat{H} \right)/Z,
\end{equation}
where $Z$ is the partition function.
In the following, it is convenient to work in the Hamiltonian eigenbasis $\{ \ket{n} \}$.
We want to show that Davies maps are characterised by the following jump operators:
\begin{equation}
\begin{aligned}
        &\hat{L}^{(1)}_{nm}(\beta) = \underbrace{\gamma\left( 1 \mp f^\pm(\beta, h_m-h_n)\right)^{1/2}}_{\alpha_{nm}^{(1)}} \ket{n}\bra{m} \\
        &\hat{L}^{(2)}_{nm}(\beta) = \underbrace{\gamma\left(f^\pm(\beta,h_m-h_n)\right)^{1/2}}_{\alpha_{nm}^{(2)}} \ket{m}\bra{n}, \hspace{0.5cm} \text{with } m < n,
\end{aligned}
    \label{eq:davies:jumpop:fermions}
\end{equation}
where $h_n$ are the eigenvalues of the Hamiltonian, $\gamma$ is the system-environment coupling (which we set to $\gamma=1$ in all the calculations presented in the main text) and $f^\pm(\beta, h_n) = 1/( \exp(\beta h_n) \pm 1)$ are the Fermi and the Bose distribution functions.
To do so, we have to prove that $\mathcal{G} \left[ \hat{\tau}_{\beta} \right]=0$.
Since $\hat{\tau}_{\beta}$ commutes with $\hat{H}$, we need to consider only the dissipator $\mathcal{D}(\mathord{\cdot})=\sum_l \hat{L}^{\nodagger}_l(\mathord{\cdot}) \hat{L}^{\dagger}_l -\frac{1}{2} \{ \hat{L}^{\dagger}_l\hat{L}^{\nodagger}_l, (\mathord{\cdot}) \}$ and have to show that
$\mathcal{D} ( \hat{\tau}_{\beta} )= \sum_{m<n} \mathcal{D}^{(1)}_{nm} ( \hat{\tau}_{\beta} ) +  \mathcal{D}^{(2)}_{nm} ( \hat{\tau}_{\beta} ) = 0$.
Indeed, for every $(n,m)$ the dissipator contributions from the jump operators $ \hat{L}^{(1)}_{nm}$ and $ \hat{L}^{(2)}_{nm}$ cancel out:
\begin{equation}
\begin{aligned}
& \mathcal{D}^{(1)}_{nm} ( \hat{\tau}_{\beta}) +  \mathcal{D}^{(2)}_{nm} ( \hat{\tau}_{\beta} ) = & \nonumber \\
& \frac{1 \mp f^\pm(h_m-h_n)}{\mathcal{N}} e^{-\beta h_m} \left ( \ket{n}\bra{n} - \ket{m}\bra{m} \right) & \nonumber \\
& +\frac{f^\pm(h_m-h_n)}{\mathcal{N}} e^{-\beta h_n} \left ( \ket{m}\bra{m} - \ket{n}\bra{n} \right) = 0, & \nonumber
\end{aligned}
\end{equation}
where we have used that $(1 \mp f^\pm(h_m-h_n)) e^{-\beta (h_m-h_n)} = f^\pm (h_m-h_n)$, which is the thermal detailed balance condition.
Thus, the Lindbladian specified by the jump operators~\cref{eq:davies:jumpop:fermions} has the thermal state as its fixed point.

Like any Lindbladian, the Davies map can be represented by a non-Hermitian matrix via vectorisation procedure~\cite{kosloff_vectorization, RevModPhys.94.045006} as
\begin{equation}
    \begin{aligned}
       &\hat{\mathcal{G}} = -i \hat{H} \otimes  \hat{\mathds{1}} + i   \hat{\mathds{1}} \otimes \hat{H}^{\scriptscriptstyle T} + \\
       &\sum_l \hat{L}^{\nodagger}_l \otimes \left ( \hat{L}^{\dagger}_l \right ) ^{\scriptscriptstyle T} -\frac{1}{2} \hat{L}^{\dagger}_l \hat{L}^{\nodagger}_l \otimes  \hat{\mathds{1}}  - \frac{1}{2}  \hat{\mathds{1}} \otimes \left( \hat{L}^{\dagger}_l \hat{L}^{\nodagger}_l \right) ^{\scriptscriptstyle T}.
    \end{aligned}
    \label{eq:lindblad:vectorized}
\end{equation}   
As we discuss in the main text, if the Hamiltonian is non-degenerate $\hat{\mathcal{G}}$ can be recast into a block-diagonal form $\hat{\mathcal{G}} = \hat{\mathcal{G}}_p \oplus \hat{\mathcal{G}}_c$, with $\hat{\mathcal{G}}_p$ describing the system's populations and $\hat{\mathcal{G}}_c$ the system's coherences
\begin{equation}
\hat{\mathcal{G}}  
 =
  \begin{blockmatrix}
    \block[gray!20](0,0)\textrm{ \color{mygreen} \bfseries popula-}(2.5,2.5)
    \block[none](0,-0.5)\textrm{ \color{mygreen} \bfseries tions}(2.5,2.5)
    \block[mygreen!20](2.5,2.5)2^L(1,-1)
    \block[mygreen!20](0,0)2^L(1,-1)
    \block[gray!20](2.5,0)\textrm{\Large \color{mypurple} \bfseries coherences}(5,-5)
    \block[mypurple!20](5.5,1)4^L - 2^L(2,-1)
    \block[mypurple!20](0.5,-4)4^L - 2^L(2,-1) 
  \end{blockmatrix},
  \label{eq:davies:block:structure}
\end{equation}
where $L$ is the number of fermions, hard-core bosons, or spin-$1/2$ particles in the system.
Moreover, the left eigenmatrices $\hat{l}_k$ associated with the system's populations are diagonal and correspond to real eigenvalues $\lambda_k$, while those associated with the system's coherences are strictly off-diagonal and the relative eigenvalues come in complex conjugate pairs.

Instead of plugging the jump operators~\cref{eq:davies:jumpop:fermions} into the general expression for the vectorised Lindbladian~\cref{eq:lindblad:vectorized}, one can also construct the two blocks of the Davies map individually, saving significant computational resources.
For this purpose, we define the matrix $\hat{J}$ collecting all the jump operators. 
For concreteness, for two qubits it reads:
\begin{equation}
    \hat{J} = \begin{pmatrix}
0 & \alpha_{01}^{(2)} & \alpha_{02}^{(2)} & \alpha_{03}^{(2)} \\
\alpha_{10}^{(1)} & 0 & \alpha_{12}^{(2)} & \alpha_{13}^{(2)} \\
\alpha_{20}^{(1)} & \alpha_{21}^{(1)} & 0 & \alpha_{23}^{(2)} \\
\alpha_{30}^{(1)} & \alpha_{31}^{(1)} & \alpha_{32}^{(1)} & 0
\end{pmatrix}
\label{eq:jump:op:matrix}.
\end{equation}
The coefficients $\alpha^{(1)}_{nm}$ and $\alpha^{(2)}_{nm}$ are obtained from~\cref{eq:davies:jumpop:fermions}. 
It can be shown that the populations block $\hat{\mathcal{G}}_p$ reads
\begin{equation}
    \hat{\mathcal{G}}_p = \hat{J}_2 + \hat{J}_s,
\end{equation}
where $\hat{J}_2$ is obtained by squaring $\hat{J}$ elementwise and $\hat{J}_s$ is a diagonal matrix with $\hat{J}_s[k,k] = - \sum_{i=0}^{2^L-1} \left(\hat{J}[i,k] \right)^2$.

Let us now construct the coherences block $\hat{\mathcal{G}}_c$, which is diagonal in the energy eigenbasis. 
For this one has to pick all possible $(4^L-2^L)/2$ couples of columns $\hat{J}[:,n]$ and $\hat{J}[:,m]$ with $n=0,1,\dots 2^L$ and $m>n$.
Each couple of columns $(n,m)$ identifies two elements $\hat{\mathcal{G}}_c[k,k]$ and $\hat{\mathcal{G}}_c[l,l]$ which are the complex conjugate of one another.
The indices of these two elements are $k = m\cdot 2^L-m+n-1$ and $l=n\cdot 2^L+m-n$.
Their real part reads
\begin{equation}
\begin{aligned}
   &\Re(\hat{\mathcal{G}}_c[k,k]) = \Re(\hat{\mathcal{G}}_c[l,l]) = \\ & -\sum_{i=0}^{2^L-1} \left( \frac{(\hat{J}[i,n])^2}{2} + \frac{(\hat{J}[i,m])^2}{2}  \right),
\end{aligned}
\end{equation}
and their imaginary part is 
\begin{equation}
    \Im(\hat{\mathcal{G}}_c[k,k]) = \Im(\hat{\mathcal{G}}^*_c[l,l]) = i(h_n-h_m) .
\end{equation}

\section{\label{app:metropolis} Metropolis algorithm for the  Mpemba effect }
\begin{figure}[t]
        \centering
        \includegraphics[width=0.5\textwidth]{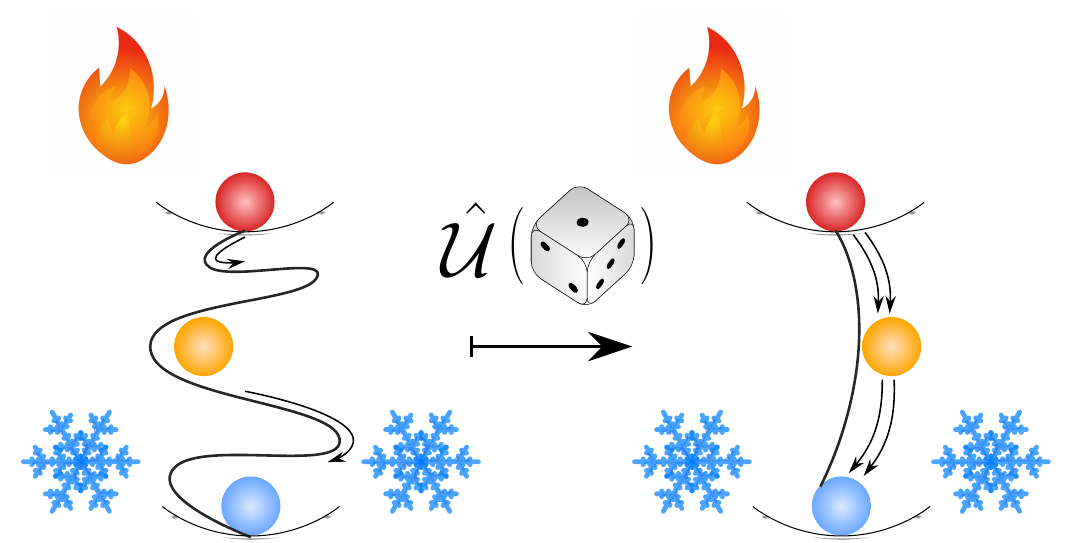}
    \caption
 {A unitary transformation obtained from a stochastic Metropolis algorithm can exponentially accelerate a relaxation process.
        \label{fig:mpemba:cartoon}
 }
\end{figure}
In this article, we have focused on Davies maps and shown how exponential speedups related to a genuine quantum Mpemba effect can be obtained for thermalisation processes.
As demonstrated by the vast recent literature~\cite{PhysRevB.100.125102, carollo_prl, bao2022accelerating, PhysRevResearch.3.043108, carollo_numerical, PhysRevE.108.014130, PhysRevLett.131.080402, chatterjee2023multiple, wang2024mpemba, caceffo2024entangled, strachan2024nonmarkovian,shapira2024mpemba, strong_mpemba_observation}, however, transformations of initial states that lead to anomalous relaxations in general Lindbladians are of great relevance.
Here we introduce a numerical method that generalises \cite{carollo_prl, carollo_numerical} to find (possibly multiple) exponential speedups for \textit{mixed} states evolving with arbitrary Lindbladians.

Given an initial state $\hat{\rho}$ and $K$ left eigenmodes, the goal is to find the unitary transformation $\hat{U}: \: \: \hat{\rho} \to \hat{\rho}'$, such that the cost function
\begin{equation}
C = \sum_{k=2}^{K+1}\abs{\Tr(\hat{l}_k \hat{\rho}')} 
\label{eq:cost:function}
\end{equation}
is minimised.
A general global unitary transformation acting on $L$ qubits is characterised by $4^L$ real parameters.
Thus, in order to be able to consider large systems, we assume that the global unitary can be decomposed in terms of single-qubit unitaries:
\begin{equation}    
    \begin{aligned}
    &\hat U = \prod_{j=1}^L \hat U_j \hspace{0.5 cm} \text{with} \\ 
    &\hat U_j(\alpha_j, \beta_j, \gamma_j, \delta_j) = \exp(i\alpha_j) \hat R_z(\beta_j) \hat R_x(\gamma_j) \hat R_z(\delta_j) ,
    \end{aligned}
    \label{eq:mpemba:unitary:Lqubits}
\end{equation}  
where $\alpha, \beta, \gamma, \delta \in [0, 2\pi]$.
With this ansatz, an optimal $\hat U$ can be obtained by performing a Metropolis search~\cite{Metropolis_original, numerical_recepies} consisting of the following steps:
\begin{enumerate}
    \item Initialise a random $\hat U$ and an effective temperature $T_{\text{eff}}=1$, rotate $\hat \rho$ with $\hat U$ and compute the cost function $C$.
    \item Pick randomly which qubit $j$ to optimise.
    \item For the selected qubit, choose stochastically which parameter $x_j \in \{ \alpha_j, \beta_j, \gamma_j, \delta_j \}$ of $\hat{U}_j$ to optimise. 
    \item Vary the selected parameter $x_j \to x_j' = x_j + \delta x_j$ by a random increment $\delta x_j \in [0, 2\pi]$, update $\hat{U}$ and $\hat{\rho}' \equiv \hat{U} \hat{\rho} \hat{U}^\dagger $ and reevaluate the cost function~\cref{eq:cost:function} with $\hat{\rho}'$ .
    \item If $C' < C$, accept the new unitary $\hat{U'} \to \hat{U}$ and update $C' \to C$. Otherwise accept the new unitary with probability $p = \exp \left( - \frac{C' - C}{T_{\text{eff}}} \right)$.
    \item If the new unitary is accepted, decrease the effective temperature by a cooling constant $\tau$: $T_{\text{eff}} \to \tau T_{\text{eff}}$.
    \item Repeat steps 4-6 $n$ times (nano iterations: optimising over one parameter). 
    \item Repeat steps 3-6 $m$ times (micro iterations: optimising over all parameters of a single qubit $j$). 
    \item Repeat steps 2-6 $LM$ times (macro iterations: optimising over all qubits).
\end{enumerate}
The algorithm is terminated once the cost function has decreased below a threshold $\epsilon$.
We dubbed this algorithm \textit{unitary Metropolis}.
The generalisation to fermionic systems is straightforward and consists of taking into account the fermionic anticommutation relations. 
This can be achieved, for instance, by replacing the single-qubit unitary $\mathds{1} \otimes \mathds{1}_2 \dots \otimes \hat{U}_j \otimes \mathds{1}_{j+1} \dots \mathds{1}_{L}$ in~\cref{eq:mpemba:unitary:Lqubits} with $\mathds{1}_1 \otimes \mathds{1}_2 \dots \otimes \hat{U}_j \otimes \sigma^z_{j+1} \dots \sigma^z_{L}$, which yields:
\begin{equation}
    \hat U^f =\prod_{j=1}^L \hat U_j \: \left ( \hat \sigma_j^z \right )^\text{mod(L-j,2)} .
    \label{eq:mpemba:unitary:Lfermions} 
\end{equation}
Note also that the same method can be applied to bosonic systems with local dimension $d$ by decomposing single-site $d$-dimensional unitary operators $\hat{U}(d)$ into two-level rotations as outlined in Algorithm 1 in~\cite{Ringbauer2022}.

The algorithm outlined above can be simplified when considering the special class of Lindbladians called Davies maps, which describe thermalisation processes.
As we discussed in~\cref{app:Davies-Lindblad}, Davies maps have a block diagonal structure indicating that populations and the coherences of an initial state evolve independently from one another.
Thus, when the initial state is diagonal (incoherent) in the energy eigenbasis, the unitary optimization can be substituted by a simpler swap of the state's populations in the following way:
\begin{enumerate}
    \item Recast the diagonal part of the incoherent $(2^L \times 2^L)$ initial state $\hat{\rho}$ into a $2^L$-dimensional vector and do the same for the targeted diagonal left eigenmatrices $\hat{l}_k$. We will indicate the vectorised matrices with double brackets $\hat{\cdot} \to || \cdot \rangle \rangle$. 
    \item Evaluate the cost function $C \equiv \sum_{k=1}^{K+1}\abs{ \langle \langle l_k | \rho \rangle \rangle }$ and initialise an effective temperature $T_{\text{eff}}=1$.
    \item Randomly select four integers $n_1, n_2, n_3, n_4 \in [0,d]$ representing the indices of four entries of $||\rho\rangle \rangle$ and perform a random permutation $P\left(n_1, n_2, n_3, n_4 \right) = \left(\tilde{n}_1, \tilde{n}_2, \tilde{n}_3, \tilde{n}_4 \right)$, excluding the trivial permutation $\left(\tilde{n}_1, \tilde{n}_2, \tilde{n}_3, \tilde{n}_4 \right) = \left( n_1, n_2, n_3, n_4 \right )$.
    \item Compute $||\rho'\rangle \rangle$ by swapping the four randomly selected elements of $||\rho\rangle \rangle$ according to the random permutation $P$ and compute the cost function $C'$ with the updated state.
    \item If $C' < C$, accept the new state $||\rho'\rangle \rangle \to ||\rho\rangle \rangle$ and update $C' \to C$. Otherwise accept the new state with probability $p = \exp \left( - \frac{C' - C}{T_{\text{eff}}} \right)$.
    \item If the new state is accepted, decrease the effective temperature by a cooling constant $\tau$: $T_{\text{eff}} \to \tau T_{\text{eff}}$.
    \item Repeat steps $3-6$ until the cost function is reduced below a threshold $\epsilon$. 
\end{enumerate}
We refer to this method as \textit{swap Metropolis}.
The same method can be applied to a classical Markovian thermalisation process, the only difference being that the eigenmodes and initial states are already vectors, so step 1. can be skipped.
We note that a swap minimisation method (which was not utilised in this work) was imple- mented in Mathematica \cite{wolfram_MaximizeOverPermutations}.
\begin{figure}[h]
    \centering
    \includegraphics[width=0.5\textwidth]{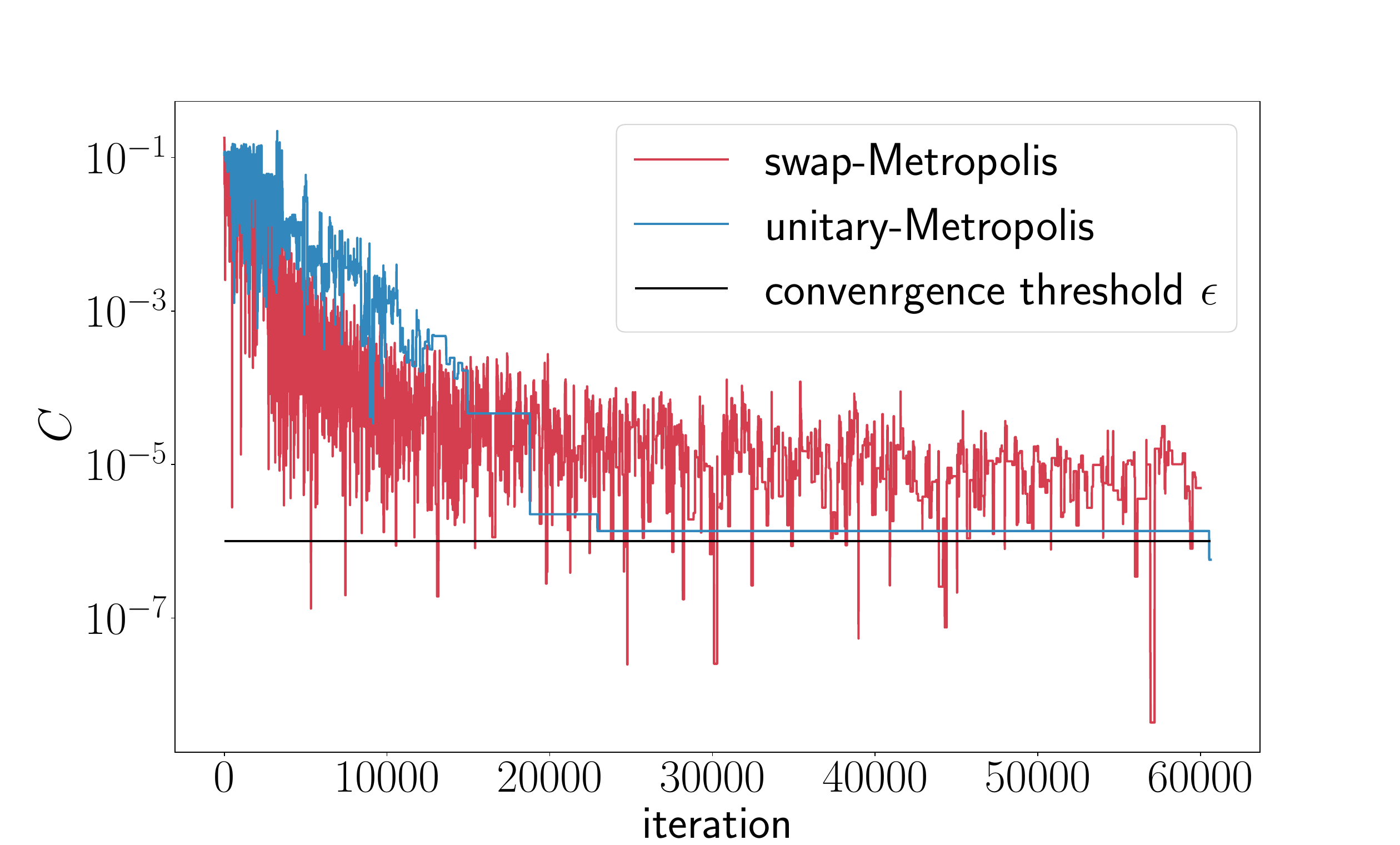}
    \caption{
 The swap Metropolis and the unitary Metropolis methods applied to the thermalisation of a \gls{TFIM} with $L=5$ and $h=J$.
 The swap Metropolis (red line) is used to decrease the overlap of a thermal state with one diagonal eigenmode with a cooling constant $\tau = 0.998$.
 The unitary Metropolis is applied to reduce the overlap of a random state with two off-diagonal eigenmodes with $\tau = 0.999$.
 We set the maximally allowed number of nano iterations to $n=200$, micro iterations to $m=20$, and macro iterations to $M=20$.}
    \label{fig:metropolis}
\end{figure}
\begin{figure}[t]
\label{subfig:swap:L1}
\includegraphics[width=0.49\textwidth]{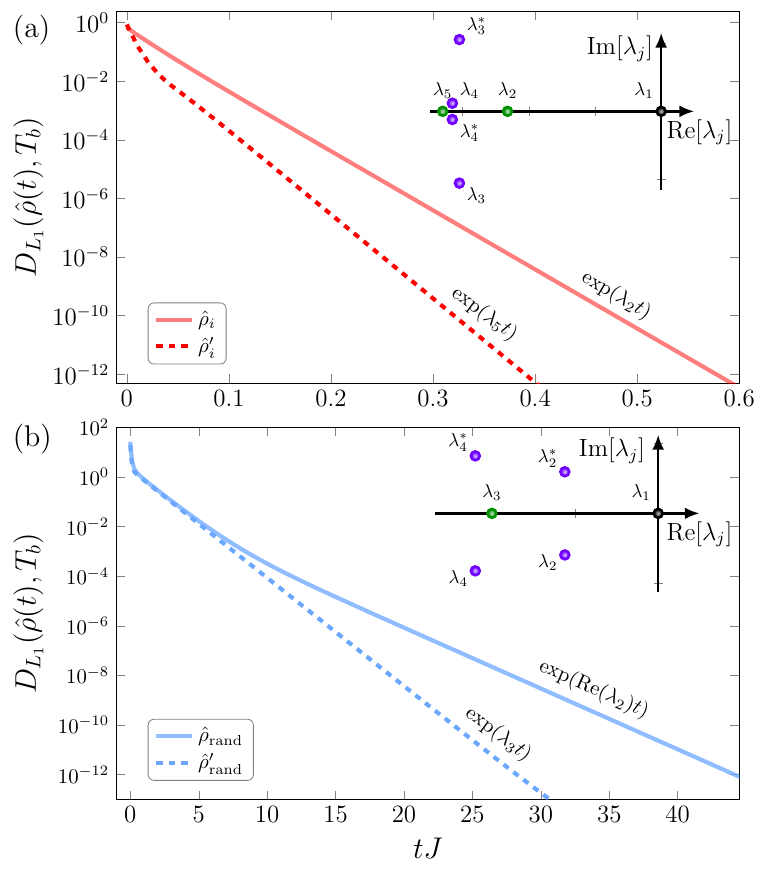}
    
    \caption
 {Exponentially accelerated thermalisations after the Metropolis optimisation ( see~\cref{fig:metropolis}).
 Panel (a): the $L_1$-distance from the thermal steady state before (full line) and after (dashed line) applying the swap metropolis optimisation during the heating process of a thermal state.
 Panel (b): the $L_1$-distance from the thermal steady state before (full line) and after (dashed line) applying the unitary metropolis optimisation during the thermalisation process of a random mixed state.
 In both cases, the Metropolis algorithms yield exponential speedups.
    \label{fig:metropolis:l1}
 }
\end{figure}

We now apply the two Metropolis algorithms to study the thermalisation of a \acrfull{TFIM}  with open boundary conditions $\hat{H} = -J \sum_{j = 1}^{L-1} \hat{\sigma}^z_j \hat{\sigma}^z_{j+1} + h   \sum_{j = 1}^{L} \hat{\sigma}^x_j$ and set $h = J$ and $L=5$.
First, we consider the heating of a thermal state from $T_i=1 \,J$ to $T_b=4 \,J$ and use the swap Metropolis algorithm to minimise the overlap of the initial state with a single eigenmode $\hat{l}_2$. 
Then, we study the thermalisation of a random mixed state (obtained by averaging over 1000 random pure states) with $T_b=0.1 \,J$ and apply the unitary Metropolis to target two modes, $\hat{l}_2$ and $\hat{l}_3$.
\cref{fig:metropolis} shows that the swap Metropolis already decreases $C$ below $\epsilon = 10^{-6}$ after around $5300$ iterations, while the unitary Metropolis crosses the convergence threshold after about $60500$ iterations.
We stress that despite the relatively large number of iterations used, both methods have low computational costs. The bottleneck for the investigation of the Mpemba effect is constituted by the diagonalisation of the quantum or classical Liouvillian and not by the minimisation algorithms.
In~\cref{fig:metropolis:l1} we show the dynamics of the $L_1$-distance from the steady state before and after the swap Metropolis and the unitary Metropolis optimisations.
In both cases, it can be seen that exponential speedups are found.
In particular, panel (a) shows that since the swap Metropolis algorithm preserves the diagonal structure of the thermal initial state, the transformed state remains orthogonal to the off-diagonal modes and the thermalisation rate is boosted from $\lambda_2$ to $\lambda_5$.

\section{\label{app:davies:applications} Two physical scenarios described by the Davies map}
\begin{figure}
    \centering
    \subfloat[\label{fig:app:tla:photons}]{
        \includegraphics[width=0.48\textwidth]{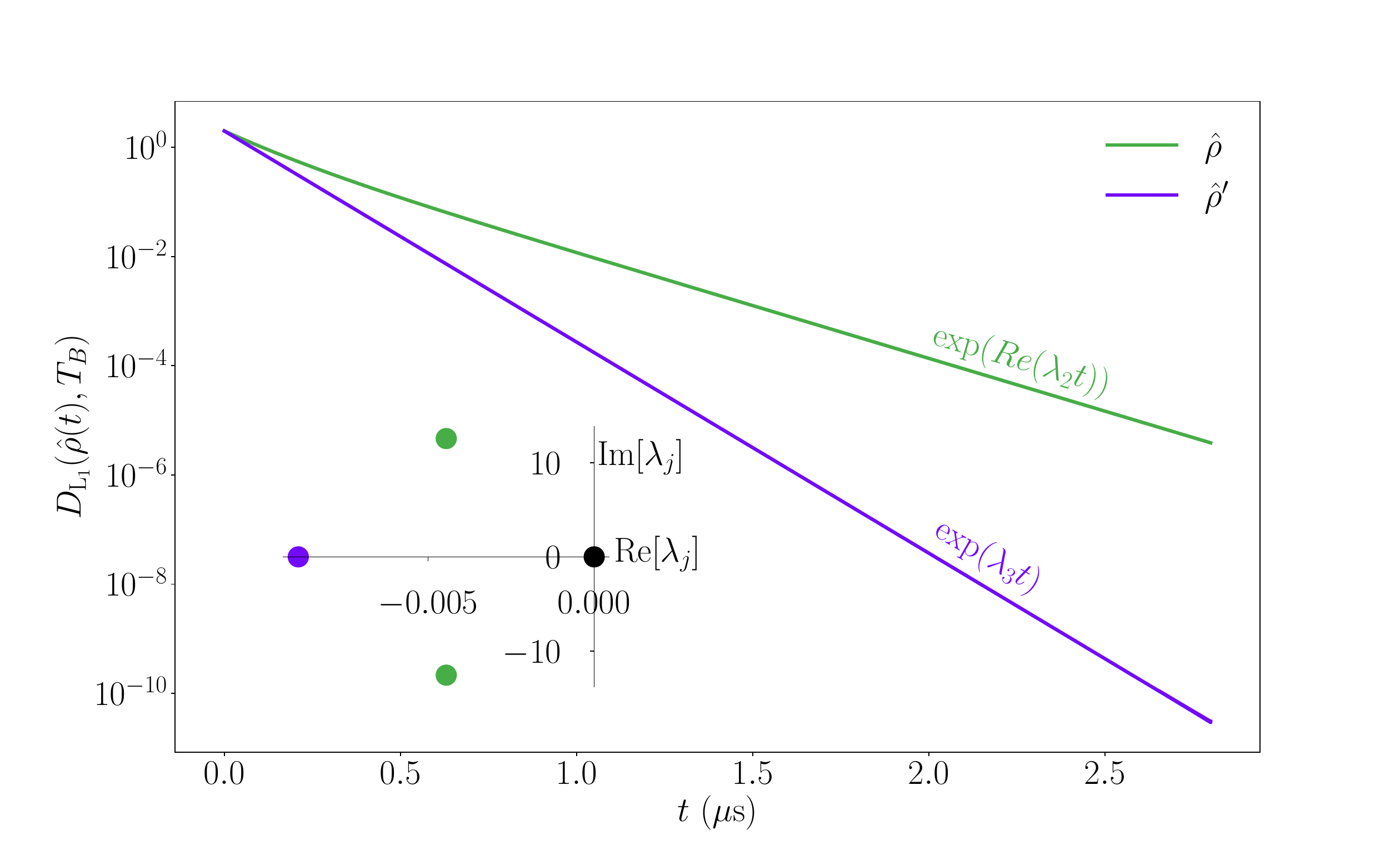}
 }
    \vspace*{-2em}
    \subfloat[\label{fig:app:dot}]{
        \includegraphics[width=0.48\textwidth]{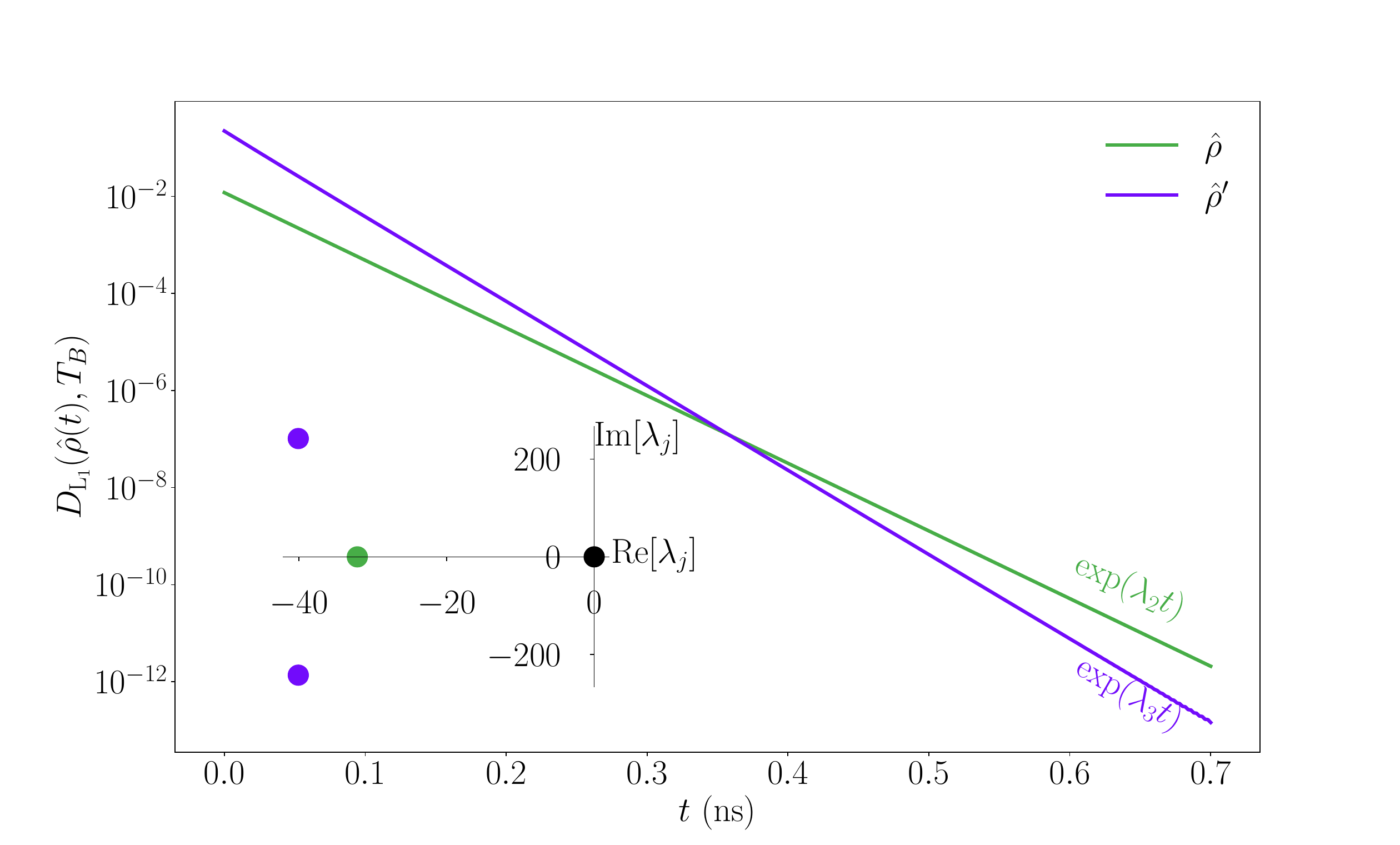}
 }
    \caption{Exponentially accelerating the thermalisation of a two-level atom in a photonic environment and of a spinful fermionic quantum dot.
 (a) The two-level atom is initialized in state $\hat{\rho}(t=0) = \ket{+}\bra{+}$ with $\ket{+} = 1/\sqrt{2}(\ket{0}+\ket{1})$. 
 Since the eigenvalue $\lambda_2$ of the Davies map is complex, the state $\hat{\rho}'(t=0)$ having zero overlap with the slowest-decaying mode $\hat{l}_2$ is obtained by applying the transformation $\hat{U}$ introduced in the main text.
 \textit{Inset}: the eigenvalues of the Davies map.  
 (b) The spinful fermionic quantum dot is initialized in the thermal state at $T_i= 0.1 \mathrm{K}$.
 For this system the eigenvalue $\lambda_2$ of the Davies map is real, and thus a transformed state $\hat{\rho}'(t=0)$ with overlap $\mathcal{O}= \abs{\Tr(\hat{l}_2 \hat{\rho}')} + \abs{\Tr(\hat{l}^\dagger_2 \hat{\rho}')} <2 \cdot 10^{-5}$ is obtained using the Metropolis-based algorithm introduced in~\cref{app:metropolis}.
 \textit{Inset}: the four eigenvalues of the Davies map with the largest real part.
    \label{fig:app:two:davies:examples}
 }
\end{figure}
\discussive{Davies maps describe the general scenario of the relaxation of quantum systems weakly coupled to Markovian heat baths.
They were recently used to model noisy quantum gates in superconducting architectures~\cite{papič2023fastestimationphysicalerror} and to study the quantum coherent contribution to entropy production in an optical experiment~\cite{xue2024evidence}.}

\discussive{For concreteness}, in this appendix, we consider two physical setups described by the Davies map to which our description of the quantum Mpemba effect can be applied.
\paragraph{A two-level atom in a photonic environment}
First, we consider a two-level atom interacting with a photonic environment.
This system is characterized by the Jaynes-Cummings Hamiltonian that, in the rotating wave approximation, reads~\cite{paule2018thermodynamics}:
\begin{equation}
    \begin{split}
    \hat{H} =& \hat{H}_\mathrm{s} + \hat{H}_\mathrm{e} + \hat{H}_\mathrm{int} = \epsilon \hat{\sigma}^+ \hat{\sigma}^- + \sum_k \omega_k  \hat{a}^\dagger_k \hat{a}_k 
    \\
    &+\sum_k g \left( \hat{\sigma}^-\hat{a}^\dagger_k e^{-i(\epsilon -\omega_k)t}+ \hat{\sigma}^+\hat{a}_k e^{+i(\epsilon -\omega_k)t}\right) ,
    \end{split}
\end{equation}
where we have set $\hbar=1$.
Here $\epsilon$ is the energy difference between the two states of the atom, $\hat{\sigma}^+$ and $\hat{\sigma}^-$ are the ladder operators for the two-level atom, $\hat{a}^\dagger$ and $\hat{a}$ are the bosonic creation and annihilation operators, respectively, and the homogeneous coupling $g$ describes the interaction strength between the two-level atom and the bosonic modes.
We now consider the photons $\hat{H}_\mathrm{e}$ to be in thermal equilibrium at inverse temperature $\beta$.
Then, following the standard steps~\cite{BRE02} (Born, Markov, and secular approximation), a Lindblad equation of Davies type for the two-level atom can be derived~\cite{paule2018thermodynamics}
\begin{equation}
    \begin{split}
    \frac{\mathrm{d}\hat{\rho}_\mathrm{s}(t)}{\mathrm{d}t} &= -i [\hat{H}_\mathrm{s}, \hat{\rho}(t)] +\gamma n_\beta( \hat{\sigma}^+ \hat{\rho}(t)\hat{\sigma}^- -\frac{1}{2} \{ \hat{\sigma}^-\hat{\sigma}^+, \hat{\rho}(t) \} ) \\
    & + \gamma(n_\beta+1)( \hat{\sigma}^-\hat{\rho}(t)\hat{\sigma}^+ -\frac{1}{2} \{ \hat{\sigma}^+\hat{\sigma}^-, \hat{\rho}(t) \} ) ,
    \end{split}
    \label{eq:app:davies:tla}
\end{equation}
with $n_\beta$ being the occupation of the two-level atom at the inverse temperature $\beta$ and $\gamma \propto g$ the damping coefficient. 
A similar approach considering a two-level system and a resonator was studied in~\cite{PhysRevA.106.042605} for a transmon qubit. 
The parameters used in~\cite{PhysRevA.106.042605} were $\epsilon=2\pi \times 4 \mathrm{GHz}$, $\gamma=2\pi \times 1.41 \mathrm{MHz}$, and $T_b=0.1\mathrm{K} = 2.08 \mathrm{GHz} \gg \gamma $.
In~\cref{fig:app:tla:photons} we show that the thermalisation of the two-level atom obeying~\cref{eq:app:davies:tla} can be exponentially accelerated by applying the exact unitary transformation outlined in the main text.
\paragraph{A spinful fermionic quantum dot}
As a second example, we study a spinful fermionic quantum dot interacting with a fermionic reservoir.
This setup can be described by~\cite{Josefsson2018-me}
\begin{equation}
    \begin{split}
        \hat{H} =& \; \hat{H}_\mathrm{s} + \hat{H}_\mathrm{e} + \hat{H}_\mathrm{int} = \sum_{\sigma=\uparrow,\downarrow} \epsilon \hat{n}_\sigma + E_C \hat{n}_\uparrow \hat{n}_\downarrow \\
        &+ \sum_{k,\sigma} \omega_{k} \hat{c}^\dagger_{k,\sigma}\hat{c}_{k,\sigma} + \sum_{k,\sigma} g \hat{d}^\dagger_\sigma \hat{c}_{k,\sigma} +\mathrm{h.c.},
    \end{split}    
    \label{eq:app:davies:tls}
\end{equation}
where $\epsilon$ is the dot's energy, $E_c$ is the Coulomb repulsion, $\omega_k$ is the energy of the reservoir-fermions, $g$ describes the coupling between the dot and the reservoir-fermions, $\hat{d}$ ($\hat{d}^\dagger$) and $\hat{c}$ ($\hat{c}^\dagger$) are the annihilation (creation) operators for the quantum dot and the reservoir, respectively and $\hat{n}_{\uparrow(\downarrow)} = \hat{d}^\dagger_{\uparrow(\downarrow)}\hat{d}_{\uparrow(\downarrow)}$.
In analogy to~\cref{eq:app:davies:tls} considering a weak coupling $g$ and a reservoir in thermal equilibrium, one can derive the Davies map for the quantum dot~\cite{wiseman_milburn_2009}
\begin{equation}
    \begin{split}
    \frac{\mathrm{d}\hat{\rho}_\mathrm{s}(t)}{\mathrm{d}t} &= -i [\hat{H}_\mathrm{s}, \hat{\rho}(t)] +\gamma n_\beta \sum_{\sigma=\uparrow,\downarrow}( \hat{d}_\sigma^\dagger \hat{\rho}(t)\hat{d}_\sigma -\frac{1}{2} \{ \hat{d}_\sigma\hat{d}_\sigma^\dagger, \hat{\rho}(t) \} ) \\
    & + \gamma(1-n_\beta)\sum_{\sigma=\uparrow,\downarrow}( \hat{d}_\sigma\hat{\rho}(t)\hat{d}_\sigma^\dagger -\frac{1}{2} \{ \hat{d}_\sigma^\dagger\hat{d}_\sigma, \hat{\rho}(t) \} ).
    \end{split}
    \label{eq:app:davies:dot}
\end{equation}
Following the experimental parameters outlined in~\cite{PhysRevA.106.042605} we set $E_c=1189\mathrm{GHz}$, $\gamma=1\mathrm{GHz}$, $T_b=2\mathrm{K} = 41.67\mathrm{GHz} \gg \gamma$. 
The dot's energy can be tuned by applying a voltage and we set it to $\epsilon = 242\mathrm{GHz}$.
In~\cref{fig:app:dot} we illustrate the exponential acceleration of the thermalisation of the spinful quantum dot dictated by~\cref{eq:app:davies:dot}.
In this case, since the eigenvalue $\lambda_2$ is real, the speedup is obtained via the numerical transformation outlined in~\cref{app:metropolis}.
Moreover, since the difference between $\Re(\lambda_2)$ and $\Re(\lambda_3)$ is small, the obtained speedup is less significant than for the two-level atom (see~\cref{fig:app:tla:photons}) and the spin chain considered in the main text.
\end{document}